\pdfoutput=1

\documentclass[times,twocolumn,final,authoryear]{elsarticle}

\usepackage{jasr}
\usepackage{framed,multirow}
\usepackage[yyyymmdd,hhmmss]{datetime}
\usepackage{amssymb}
\usepackage{latexsym}
\usepackage{scalerel,stackengine}
\stackMath
\newcommand\reallywidehat[1]{%
\savestack{\tmpbox}{\stretchto{%
  \scaleto{%
    \scalerel*[\widthof{\ensuremath{#1}}]{\kern.1pt\mathchar"0362\kern.1pt}%
    {\rule{0ex}{\textheight}}
  }{\textheight}%
}{2.4ex}}%
\stackon[-6.9pt]{#1}{\tmpbox}%
}
\usepackage{textcomp}

\usepackage{url}
\usepackage{xcolor}
\usepackage{amsmath,amssymb}
\definecolor{newcolor}{rgb}{.8,.349,.1}

\usepackage[citebordercolor=white]{hyperref}

\journal{Advances in Space Research}

\begin{document}

\verso{C. Tiburzi \textit{et. al}}

\begin{frontmatter}

\title{Validation of heliospheric modeling algorithms through pulsar observations I: Interplanetary scintillation-based tomography}

\author[1]{C. Tiburzi\corref{cor1}}
\ead{tiburzi@astron.nl}
\cortext[cor1]{Corresponding author
}
\author[2]{B. V. Jackson}
\author[2]{L. Cota}
\author[3]{G. M. Shaifullah}
\author[1]{R. A. Fallows}
\author[4]{M. Tokumaru}
\author[1]{P. Zucca}

\address[1]{ASTRON $-$ the Netherlands Institute for Radio Astronomy, Oude Hoogeveensedijk 4, 7991 PD Dwingeloo, The Netherlands}
\address[2]{Center for Astrophysics and Space Sciences, University of California, San Diego, LaJolla, California, 92093-0424, USA}
\address[3]{Dipartimento di Fisica ``G. Occhialini'', Universit\`a di Milano-Bicocca, Piazza della Scienza 3, 20126 Milano, Italy}
\address[4]{Institute for Space-Earth Environmental Research, Nagoya University, Furo-cho, Chikusa-ku, Nagoya 464-8601, Japan}

\received{\today}
\finalform{\today}
\accepted{\today}
\availableonline{\today}
\communicated{X. XXXXXX}

\begin{abstract}
Solar-wind 3-D reconstruction tomography based on interplanetary scintillation (IPS) studies provides fundamental information for space-weather forecasting models, and gives the possibility to determine heliospheric column densities. Here we compare the time series of Solar-wind column densities derived from long-term observations of pulsars, and the Solar-wind reconstruction provided by the UCSD IPS tomography. This work represents a completely independent comparison and validation of these techniques to provide this measurement, and it strengthens confidence in the use of both in space-weather analyses applications.
\end{abstract}


\end{frontmatter}


\section{Introduction}\label{sec:intro}

\begin{figure}
    \centering
    \includegraphics[scale=0.4,trim={2cm 5cm 5cm 0}]{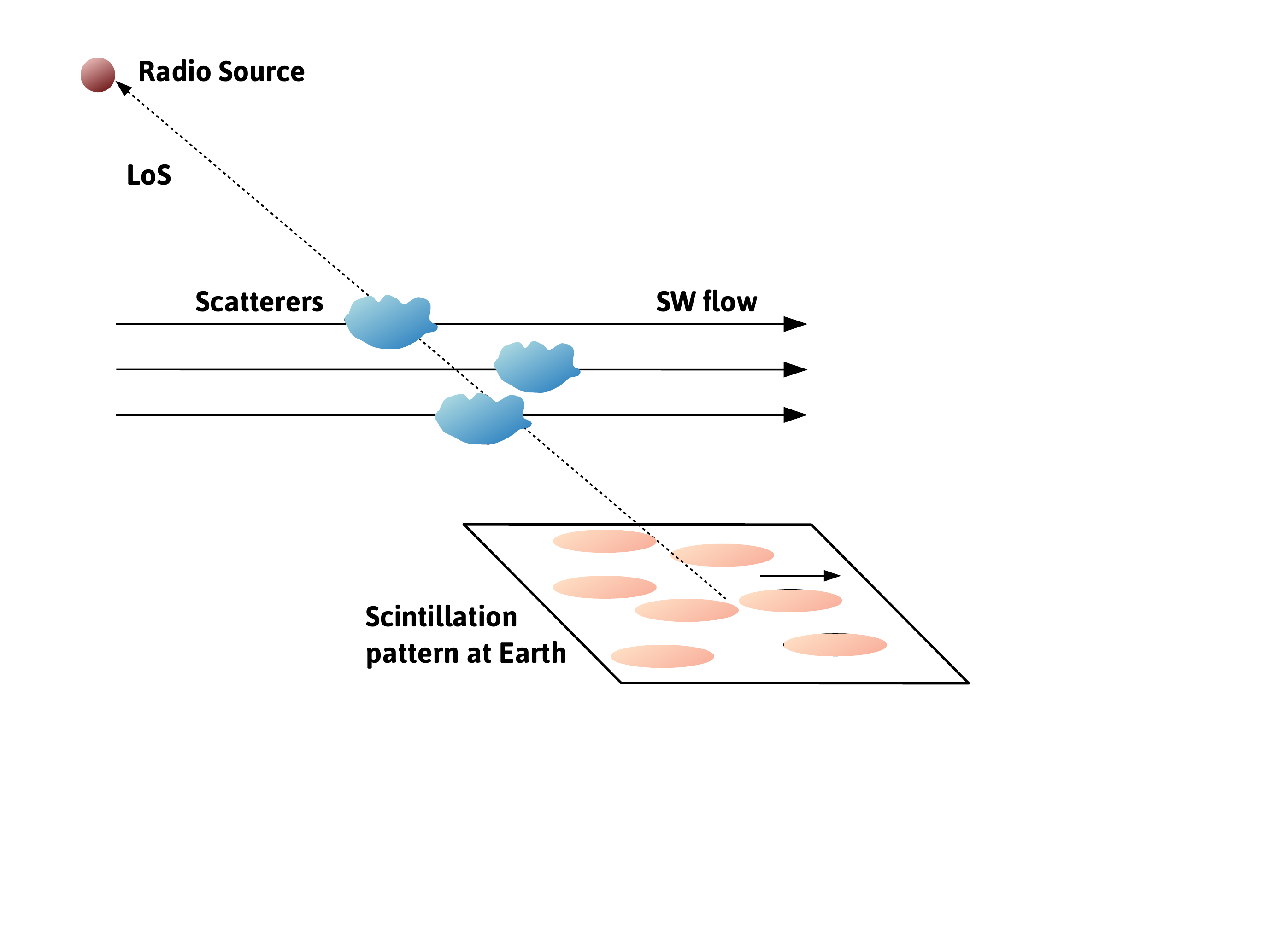}
    \caption{In intensity scintillation, scatterers along a LoS to a point-like radio source produce an intensity pattern on the surface of the Earth moving outward from the Sun at SW speed}
    \label{fig:ips}
\end{figure}

Observations of interplanetary scintillation (IPS, \citealt{Clarke:1964,hsw64,ack80}) have long been used to remotely sense small-scale ($\lesssim$500~km) density variations in the Solar wind (SW), as these variations cross the line-of-sight (LoS) to a point-like radio source. Such inhomogeneities disturb the signal from radio sources, producing intensity variations that, projected onto the Earth’s surface, form an intensity pattern travelling away from the Sun at SW speeds (see Figure~\ref{fig:ips}). In weak scintillation theory greater scintillation levels mean that more highly-variable, small-scale and, consequently, higher-density scatterers are present along the LoS. These scintillation levels can thus be used as a proxy for the SW bulk density \citep{tap86,tkf07,jhb11,jbc20}, usually determined based on a \textit{normalized scintillation level}, or g-level (g), of the IPS radio source signal in Stokes I, 
as shown in:

\begin{equation}\label{eq:glevel}
    g^2 = \frac{\langle \Delta I^2 (\epsilon) \rangle}{\reallywidehat{\langle \Delta I^2 (\epsilon) \rangle}}
\end{equation}

where $g^2$ is equal to $\langle \Delta I^2 (\epsilon) \rangle$, the power of the scintillation response averaged over a short (few-minute) time for a given observation at a particular Solar elongation $\epsilon$ (angular distance between the source and the Sun in degrees), normalized to the mean of the same quantity calculated over an annual collection of such measurements for a given radio source at the same 
elongation (see \citealt{tok00}, or Jackson et al., this issue). The g-level value 
is a proxy for the SW density fluctuations, and is also related to the bulk plasma density by iteratively matching the following equation to in-situ measurements:

\begin{equation}\label{eq:gleveltomo}
    g = A R^\alpha N^\beta
\end{equation}

where A is a constant, R is the radial distance from the Sun and N is the proton bulk density. The constant A and values of the powers $\alpha$ and $\beta$ have been determined in the past 
using the \textit{UCSD 3-D reconstruction technique} to match model densities to those measured in-situ at Earth.

The UCSD 3-D reconstruction technique fits initial model velocities perpendicular to the LoS, and g-levels to those observed by the IPS. The assumed, initial model velocity values are used to track each location on an inner source surface set at 15 Solar radii (Rs, below all lines of sight) outward to a heliospheric outer boundary, usually set at 3 AU. Each point in the volume retains knowledge of the time and spatial location of its origin on the source surface by what we call a ``traceback matrix''. This is derived from a SW model that conserves mass and mass flux from the source surface assuming radial outflow. Each position in the heliospheric volume, in addition to its initial location, holds the change of the velocity or density parameter from its initial source surface position.

In the UCSD 3-D reconstruction technique, the LoSs towards targeted radio sources are divided into segments and the SW velocity and density in each segment are calculated by using an interpolation scheme to the nearest velocity and density values determined by the traceback matrix. The model g-levels per segment are calculated with initial parameters for Equation~\ref{eq:gleveltomo}, and iterated along the LoS multiplied by the LoS weights derived by \citet{you71} assuming weak scattering, and compared to those observed. Concerning the velocity, the model values perpendicular to each LoS segment are weighted by the density and weak scattering weights to match the observed velocity values. The changes required to match the observed values are then projected back to the source surface along with the segment weights. Here, these source surface changes are then formally inverted with their weights to provide new boundary values for the SW model. Prior to providing the new model traceback matrix, these source surface Carrington maps of velocity and density are smoothed and completely filled using a combination of 2-D Gaussian spatial and temporal filters. The new model values are then projected outward to provide the next iteration fit.

We generally iterate 9 times, remove radio sources where the LoS fits for the velocity fall beyond a deviation of 3 sigma from the mean (about 1\% of the radio source number) and then iterate another 9 times. Extensive study of this process has shown that the final iterated values are insensitive to starting source surface values, and that most of the convergence occurs within one or two iterations \citep{jhk98, jbh08}.

\begin{figure}
    \centering
    \includegraphics[trim={2cm 0 0 0},scale=0.25]{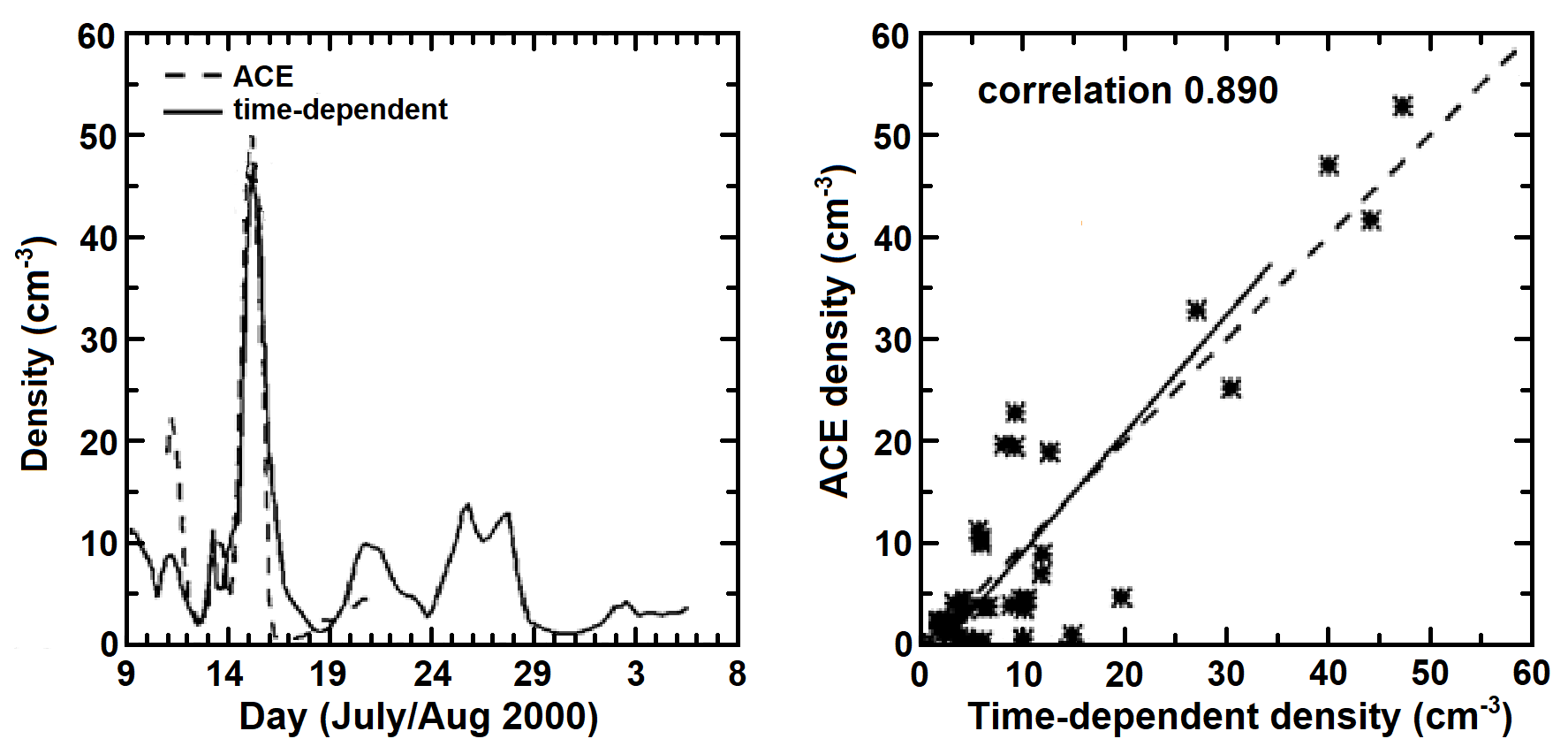}
    \caption{Time series of the IPS model compared with near-Earth Advanced Composition Explorer (ACE) level zero in-situ density over 11 days the CME event happening on the 14th of July 2000 (left panel). Power values $\alpha$ and $\beta$ were modified to provide a best match to the Pearson's R correlation (right panel).}
    \label{fig:fit}
\end{figure}

The UCSD 3-D reconstruction technique yields heliospheric structure morphology in density and velocity predictions and forecasts beyond the analysis run time for both Stream Interacting Regions (SIRs) and Coronal Mass Ejection (CME) arrival at Earth and at other locations within the volume reconstructed. Nowadays, the heliospheric parameters are well known at the locations of different spacecraft within the heliosphere while, in the past, such locations were limited to near-Earth spacecraft, close to where the observer’s LoS originate. The values of $\alpha$ and $\beta$ were originally derived in Equation~\ref{eq:gleveltomo} by matching different large variation of densities and velocities observed over many years of prior analyses using in-situ measurements of these values at Earth. In particular, they 
are derived through the maximization of the Pearson’s R correlation coefficient while comparing model velocities and densities with in-situ values (see Figure~\ref{fig:fit} for a sample of a density comparison for the Bastille Day CME event). Although the LoS segments that come close to the Sun provide some information about the value of $\alpha$ that varies the relationship of scintillation level to density with Solar distance in Equation~\ref{eq:gleveltomo}, the strongest influence in comparison with heliospheric structures is from the power $\beta$ that sets the Pearson’s R correlation relationship at Earth.

Many synopses of the 3-D reconstruction technique for space weather studies and forecasting of SIRs and CMEs can be found in the literature (e.g., \citealt{ktw98,jhk98,jbh06,jbh08,jac10,jch13}; a review of the technique and its background is in \citealt{jhb11} and references therein; and more recently in \citealt{jbc20}). The tomographic UCSD IPS 3-D reconstruction technique for near-real-time predictions used to drive the ENLIL 3-D MHD modeling (also available in near real time) are given in \citet{jch13,joy15}.

    An alternative method to derive the column density of free electrons in the SW is based on radio-frequency observations of pulsars \citep{bb69}. Pulsars are fast-rotating, highly-magnetized neutron stars, i.e., the collapsed cores of stars with an initial mass ranging between 8 and 20 Solar masses. The radio emission from pulsars is collimated in two beams roughly centered on the magnetic poles of the star, and as they sweep the space by co-rotating with it, a terrestrial observer can detect them as a periodic source of radiation. The radio emission from pulsars undergoes a series of modifications while it propagates through ionized and magnetized media, whose magnitude depends on different parameters of the traversed plasma. For this reason, pulsars are excellent probes to study the ionized interstellar medium (IISM, see \citealt{ars95}), the Solar wind \citep{cr72}, and the ionosphere \citep{pnt19}. The most widely-studied propagation effect is \textit{dispersion}, i.e., the introduction of a frequency-dependent delay in broadband radio emission determined by the amount of free electrons in the traversed plasma (see, e.g., \citealt{kcs13,jml17,dvt20}). In particular, by measuring dispersion in pulsar data it is possible to determine the column density of free electrons in the plasma, and  because of the revolution of the Earth about the Sun we can separate the individual contributions given by the IISM and the SW (the ionospheric contribution to DM has not been detected yet). 
    
    With this article, we will compare such pulsar-derived measurements to the results yielded by IPS-based tomography, and we will provide a reciprocal validation of the two techniques.

    This article is organized as follows: in Section~\ref{sec:datamethod} we will outline the details of IPS tomography, and pulsar-based derivation of free electron column density, in Section~\ref{sec:results} we will compare the results obtained with the two approaches, that we will discuss in Section~\ref{sec:discussion} before concluding in Section~\ref{sec:conclusions}.

\section{Datasets and methods}\label{sec:datamethod}

The data and methods used for the IPS 3-D density reconstructions and pulsar dispersion measurements provide a way to verify both of the techniques from two very different analysis types. The 3-D reconstructions have been used over many years in space weather applications, and while they match \textit{in-situ} measurements well at cadences of one day or better at Earth, they are a sparse data set from a few hundred to several thousand LoSs in a given month-long interval. Additionally, as stated in the introduction, these analyses are provided by a non-linear data fit to a model where average scintillation levels at distances closer to the Sun, especially in the Solar polar regions, are difficult to approximate using data obtained at 327 MHz with radio telescopes located in Toyokawa, Fuji, and Kiso (Japan) run by the Institute for Space–Earth Environmental Research (ISEE). Pulsar dispersion measures offer a more direct estimate of electron density along a given LoS, but the measurement, obtained over approximately 3-hour intervals sums to the interstellar dispersion that can vary on the order of days to months and must be estimated and separated from the heliospheric signal.

\subsection{3-D Heliospheric distributions derived from IPS tomography}\label{sec:ipstomo}

The UCSD IPS 3-D reconstructions provide a 3-D density model extending from 15 Rs out to 3 AU, and interpreted at 6-hour intervals (at 3 UT, 9UT, 15UT, and 21UT) thanks to the conversion from g-level using Equation~\ref{eq:gleveltomo}. These volumes have been fit to the LoSs that extend through them at different times and at different locations over a period of approximately 12 days; the time it takes heliospheric structures to travel outward along LoSs to the outer heliospheric boundary. Each volume presents a new visualization of the data, i.e., density ecliptic cuts, meridional cuts through Earth, and remote observer views (see the website analyses at \url{https/ips.ucsd.edu}). We limit each observational data set to a time period of one Carrington rotation in length ($\sim$27 days) to provide time dependent fits of the data for one whole Solar rotation. The Carrington rotation is actually begun one half rotation before the portion presented and up to five days beyond (for a total of 45 days) simply to provide LoSs that include data from near the Sun that may also be present to the edge of the heliosphere and measured in the observations used and presented later. The inclusion of IPS LoSs following the Carrington rotation provides a buffer to insure that LoSs at later times than the end of the Carrington rotation are also included. From each of the 6-hour volumes 
we determine the precessed RA and Dec locations of the pulsar at the current volume time and 
provide an integrated LoS measurement of density through the volume element. This gives the results shown in Section~\ref{sec:results}. 

Such LoS measurement is referred to 
proton density, therefore, the implicit assumption in these analyses (besides those of the model validity) is that the numbers of protons used to calibrate the IPS are equivalent to the numbers of electrons. Protons are used in these analyses 
because they are the most abundant heavy plasma particle and thus easiest to measure in most \textit{in-situ} plasma monitors. However, protons are not the only ion present in heliospheric plasma; helium ions are also present and they vary in abundance of from 2\% to 15\% by number at different times that are associated with different heliospheric structures along the LoS, SW speeds, and show an increase in heliographic latitude throughout the solar cycle (e.g., see \citealt{ksl07}). The general values given in these analyses is a Helium number abundance of around 4\%. The SW is electrically neutral, and so for each helium ion two electrons are present. Thus, there is a small necessary increase needed 
in our calibration scheme. Here for consistency we assume that a Helium abundance of 4\% is present and the number of electrons we need to multiply our proton values by is 8\%, with the caveat that any given LoS segment can have an associated error up to at least this amount.

\subsection{Pulsar Dispersion Measures}\label{sec:dm}
As mentioned in Section~\ref{sec:intro}, dispersion consists of a frequency-dependent delay in a propagating broadband emission introduced by free electrons in the traversed plasma. This happens because the group velocity $v_g$ is decreased by a factor represented by the refractive index $\mu$ of the medium, so that:
    
    \begin{equation}
        v_{\rm g} = c\mu = c \sqrt{1 - \left ( \frac{f_{\rm p}}{f} \right)^2 } 
    \end{equation}
    
    where $f$ is the radiation frequency, and $f_p$ is the plasma frequency, that can be written as $e^2n_e^2/(\pi~m_e)$, with $e$ being the electric charge, $n_e$ the electron density of the medium, and $m_e$ the electron mass. The delay $t$ between the arrival times of radiation with frequencies $f_1$ and $f_2$ (with $f_1>f_2$) due to dispersion can be written as:
    
    \begin{equation}\label{eq:tdisp}
    t = \frac{1}{c} \int_0^d \frac{f_{\rm p}^2}{2} ( f_2^{-2} - f_1^{-2})\,{\rm d}l = \mathcal{D}\, (f_2^{-2} - f_1^{-2})\,  \mathrm{DM}
\end{equation}

where $\mathcal{D}= e^2/(2 c\pi m_{\rm e})$ is the \textit{dispersion constant}, and the \textit{dispersion measure} is given by $\int_0^d n_{\rm e} {\rm d}l$, with $d$ being distance between the pulsar and the observer (see e.g. \citealt{lk04}). In other words, the DM parameter represents the column density of free electrons in the traversed plasma. It is possible to calculate the DM parameter from a radio-observation of a pulsar thanks to the procedure of \textit{pulsar timing} (see e.g., \citealt{lk04}), whose result is the derivation of a set of parameters (the timing model) such as its spin frequency, astrometry, the DM etc., that can describe the pulsar itself. For the purpose of pulsar timing, each pulsar observation is ``folded'' (i.e., all the received pulses are stacked in phase) and the used frequency band is divided into a customizable number of subbands (or frequency channels). After this, a template of the pulse profile (i.e., the light curve) is built (the template can or not be frequency-resolved), and used as a reference to derive the times-of-arrival (ToAs) of the pulses contained in the channels in each observation. These data-derived ToAs are then compared with the ToAs that the timing model, instead, predicts. If the model is optimal, the comparison yields white residuals. Alternatively, the residuals will show structures that can be linearly fit for the parameters of the timing model, including the DM, hence refining the model itself. Equation~\ref{eq:tdisp} shows that the DM can change in time, for example because of an intrinsic variation in $n_e$, but also, and especially, because pulsars have very high proper motions, and hence the line-of-sight (LoS) moves fast to follow them. Thus the LoS can span different parts of the IISM and the SW, each having different distributions of free electrons, causing fluctuations in the DM. Equation~\ref{eq:tdisp} also shows that the dispersion-induced delay increases quadratically with the decrease of the radiation frequency, meaning that DM can be measured more precisely in the lower part of the radio frequency band. For this reason, low-frequency observatories such as the LOw Frequency ARray (LOFAR, see \citealt{vwg13}) are preferential instruments to perform dispersion studies \citep{sha11}.

LOFAR is a low-frequency interferometer operating between 10 and 250 MHz, whose stations are located in a large number of European nations, with a higher concentration in the Netherlands. Pulsar observations with LOFAR are typically conducted between 110 and 190~MHz, and pulsar monitoring campaigns have been ongoing since 2012 with the core of LOFAR, and the individual stations located in Germany, France, and Sweden, with a weekly to monthly cadence.

\begin{figure}
    \centering
    \includegraphics[scale=0.2]{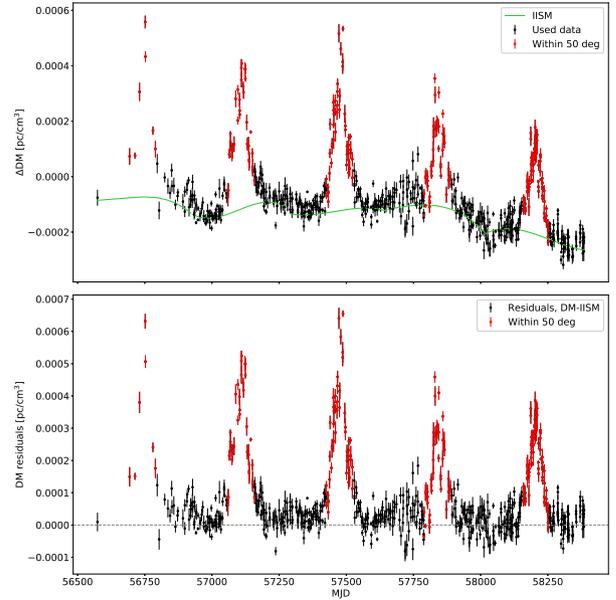}
    \caption{\textit{Top panel}: In black is reported the DM time series of PSR~J0034-0534, with the data obtained when the pulsar was at an angular distance less than 50$^\circ$ highlighted in red. The green light shows the IISM model derived by \citet{tsc21}. \textit{Bottom panel}: The same DM time series, subtracted of the IISM component.}
    \label{fig:0034}
\end{figure}

The data used in this paper were obtained in the study conducted by \citealt{tsc21} (see details about the data analysis therein). In the article, the authors calculated one value of $\Delta$DM (i.e., the difference between the DM of the timing model and the data-derived one) per observation for 43 pulsars in the aforementioned dataset, characterized by an Ecliptic latitude between $\pm20^\circ$. These DM time series were given by the sum of the IISM and the SW contribution. To study the SW only, the authors developed a Bayesian-based algorithm to disentangle the IISM part operating on 460-day-long segments of the time series. Of the initial sample, 14 pulsars were confirmed to show SW signatures, represented by a DM excess occurring during the Solar conjunction (see Figure~\ref{fig:0034}). For this article, we will use the DM time series of two pulsars, whose characteristics are reported in Table~\ref{tab:psr}: PSRs~J0034-0534 and PSR~J1022+1001.

\begin{table*}
    \centering
    \begin{tabular}{c|cccccc}
    \hline
    Pulsar & Ecliptic & RAJ & DecJ & DM$_{\rm r}$ & P0 & Analyzed \\
     & Latitude [deg] & [hh:mm:ss] & [dd:mm:ss] & [pc/cm$^3$] & [ms] & years\\
        \hline
    J0034-0534 & -8.53 & 00:34:21.8 & -05:34:36.7 & 13.795 & 1.87 & 2016 \\
    & & & & & & 2017 \\
    & & & & & & 2018 \\
    \hline
    J1022+1001 & -0.06 & 10:22:58.0 &  +10:01:52.8 & 10.252 & 16.45 & 2016 \\
     & & & & & & 2018 \\
     \hline
    \end{tabular}
    \caption{Characteristics of the pulsars analyzed in this article.}
    \label{tab:psr}
\end{table*}

\begin{figure*}
    \centering
    \begin{tabular}{cc}
     \includegraphics[scale=0.53,trim={0 0 0 0.2cm}]{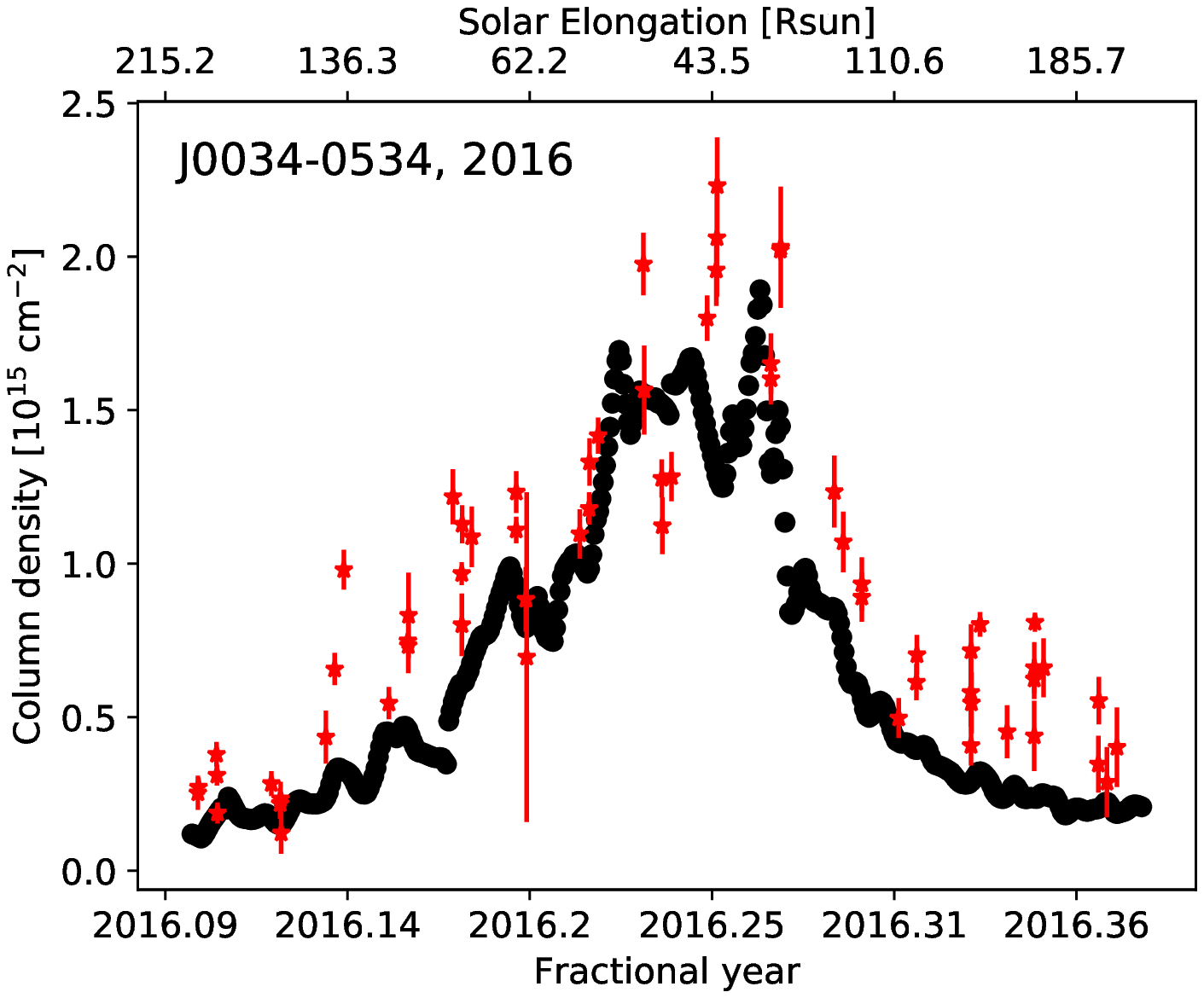}    &  \includegraphics[scale=0.53,trim={0 0 0 0.2cm}]{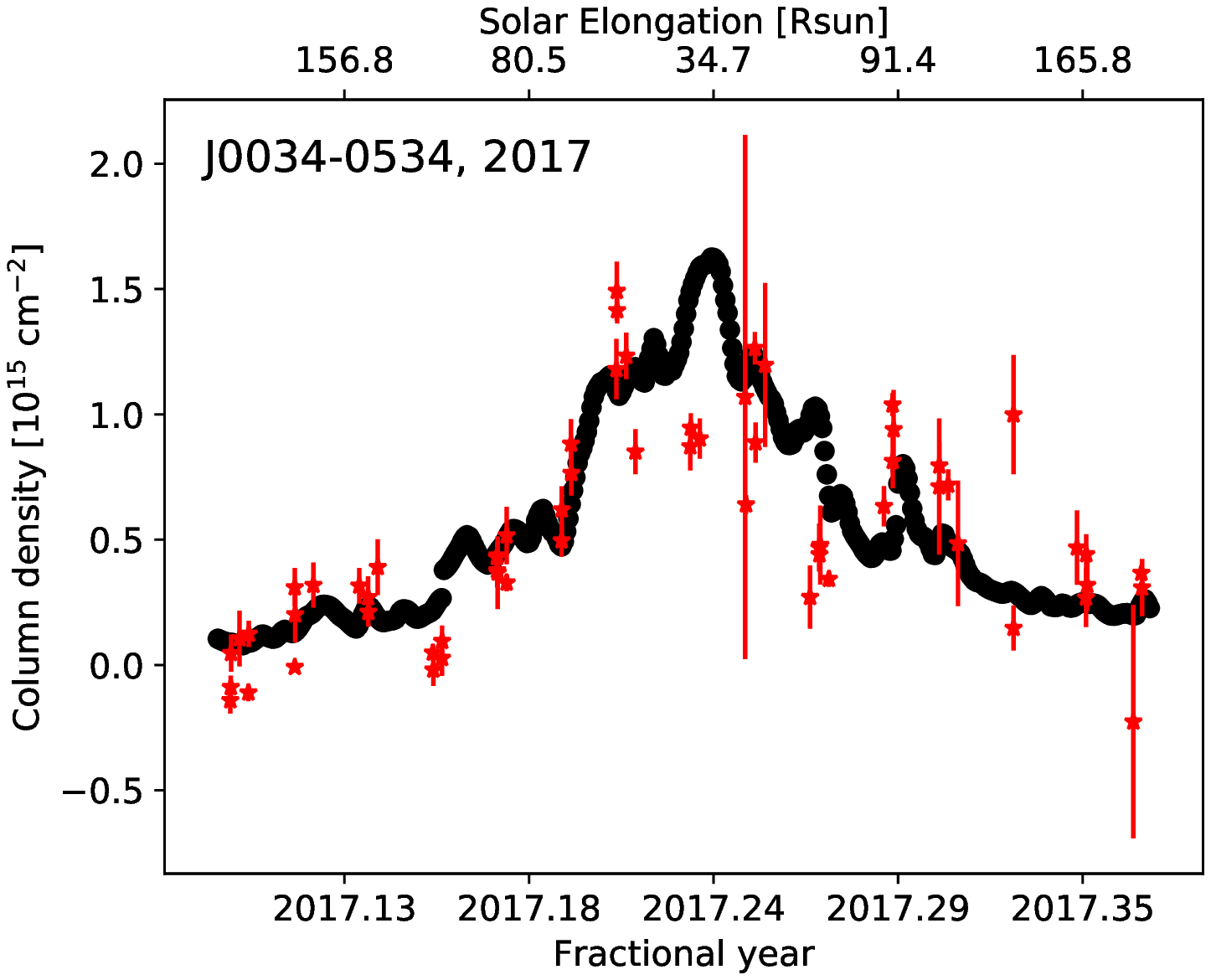}\\
     \includegraphics[scale=0.53,trim={0 0 0 0.2cm}]{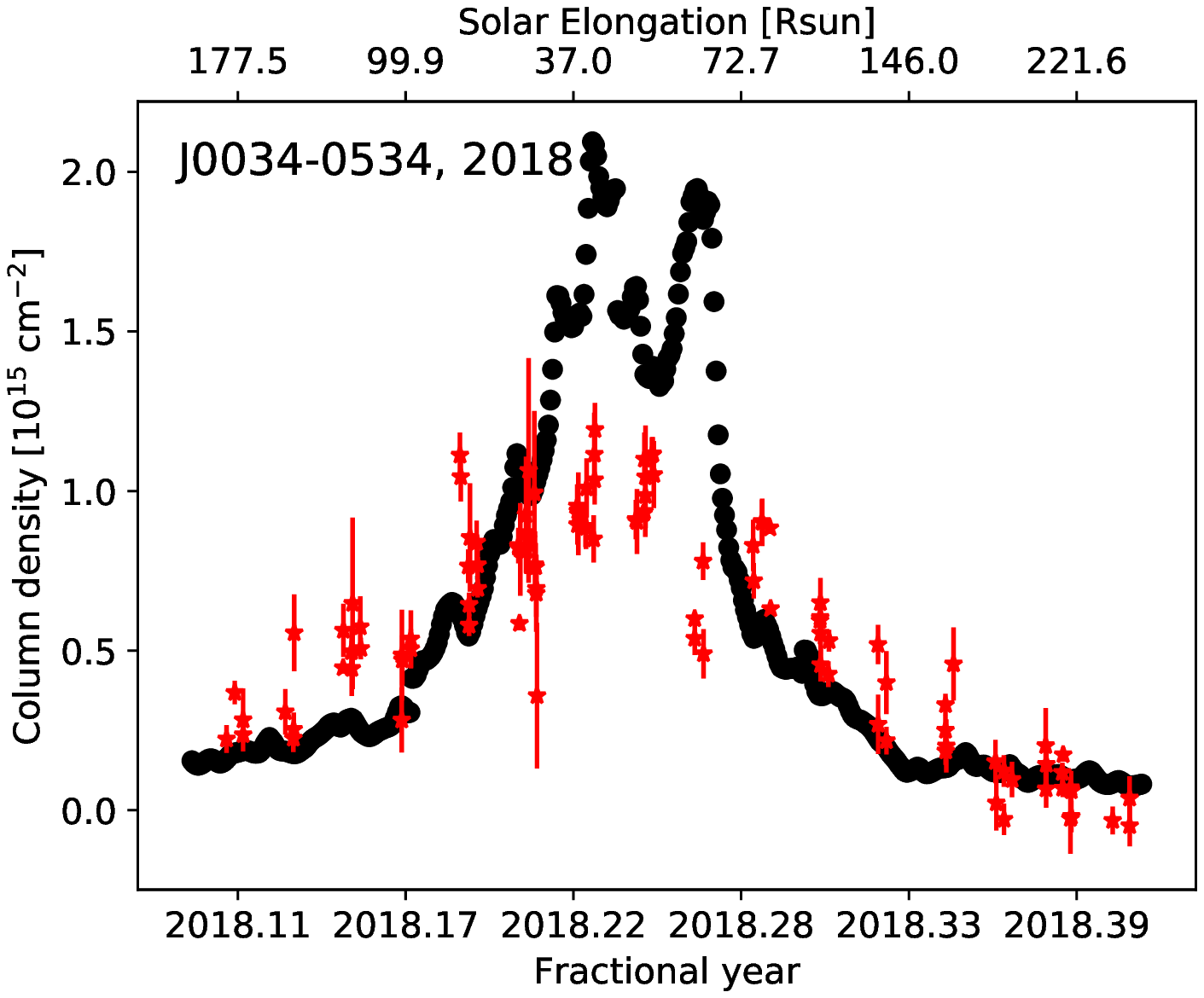}    & 
     \\
     \includegraphics[scale=0.53,trim={0 0 0 0.2cm}]{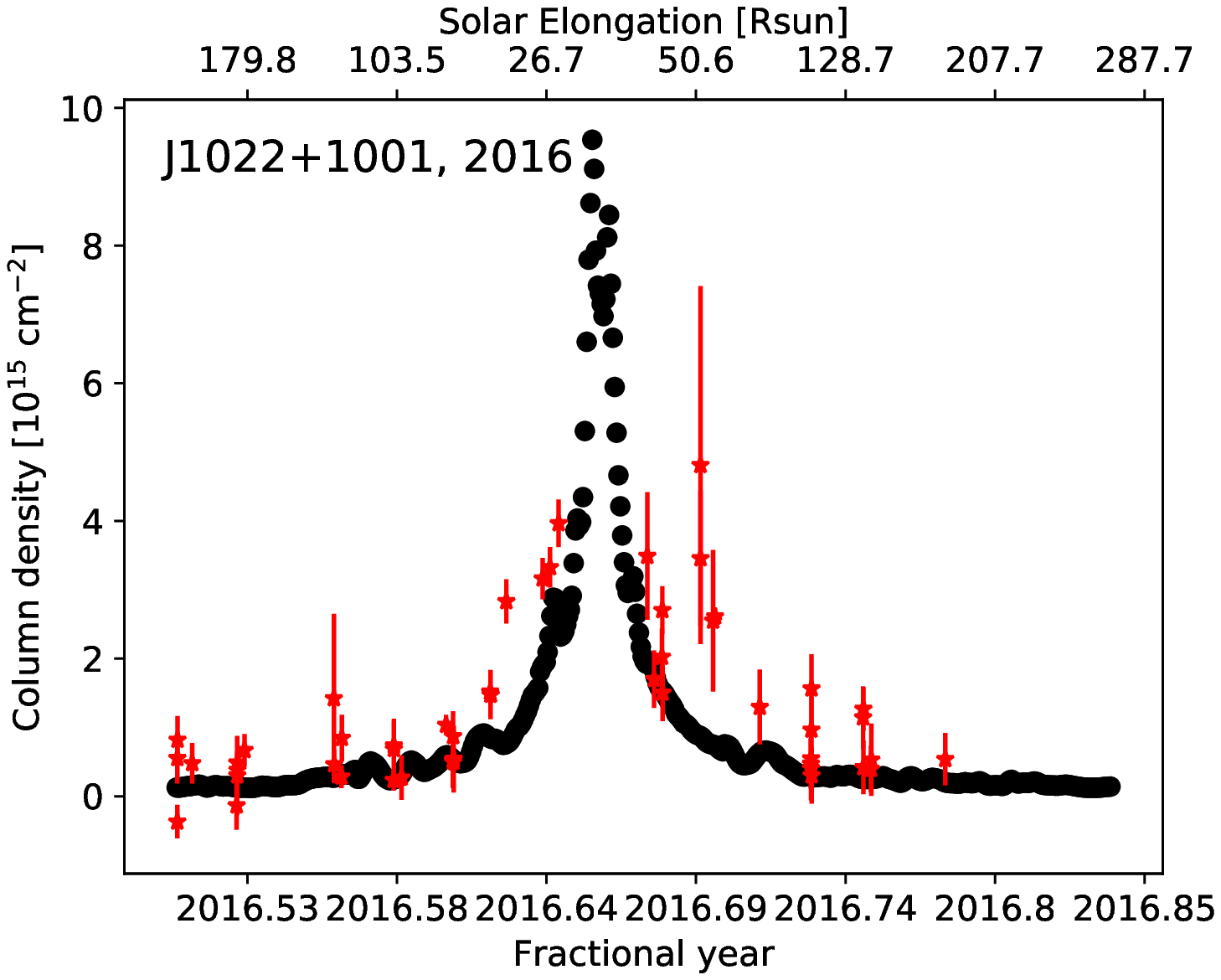} &
     \includegraphics[scale=0.53,trim={0 0 0 0.2cm}]{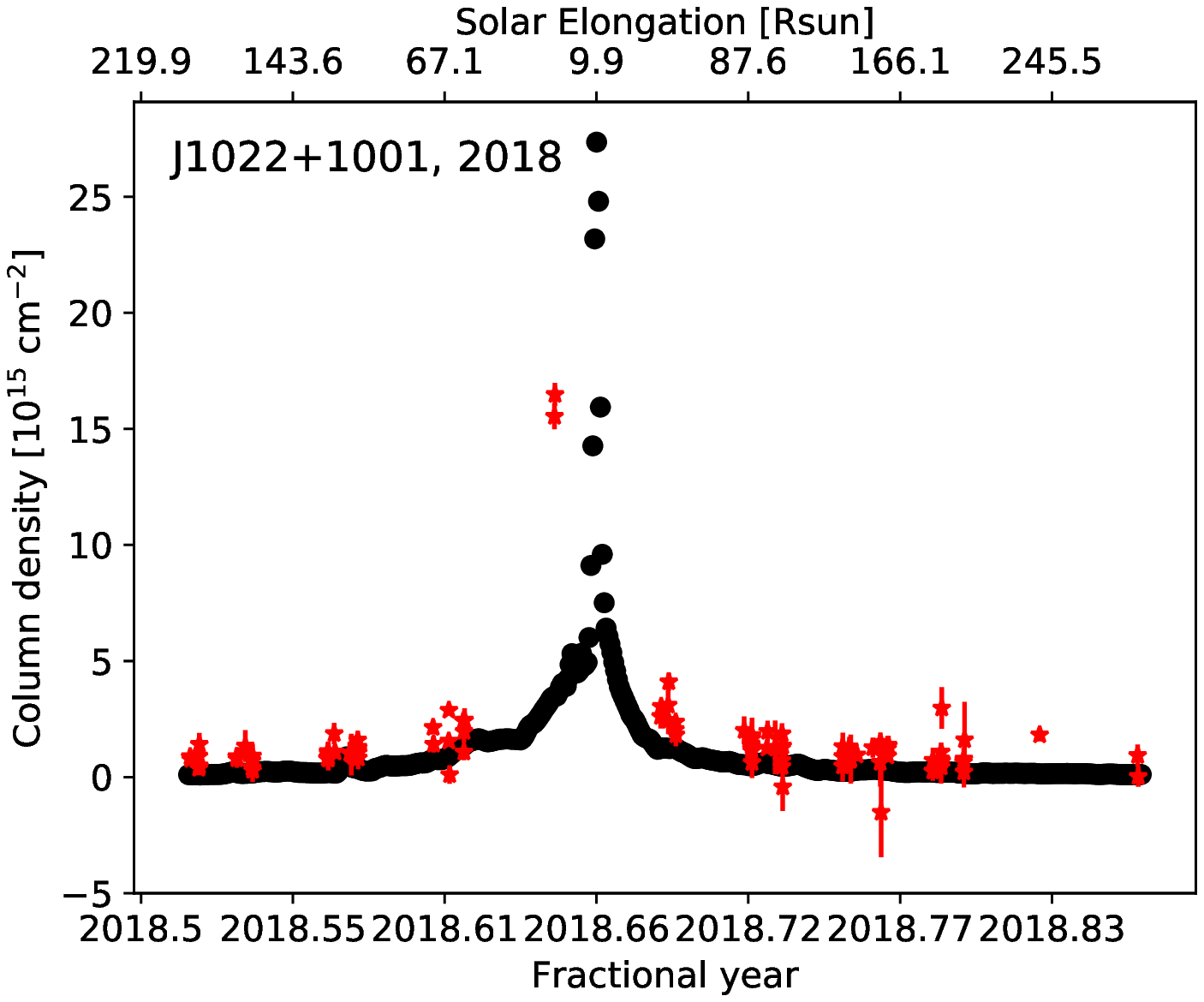}   \\
    \end{tabular}
    \caption{Comparison between the time series of electron column densities derived from the UCSD 3-D reconstruction technique (in black) and from LOFAR observations (in red). From top left to bottom right, the data are presented for: PSR~J0034-0534 in 2016, PSR~J0034-0534 in 2017, PSR~J0034-0534 in 2018, PSR~J1022+1001 in 2016, PSR~J1022+1001 in 2018.}  
    \label{fig:results}
\end{figure*}

\section{Results}\label{sec:results}

We present our comparison of the 3-D reconstruction analyses with the pulsar DM time series over intervals of about 5 months in length. These are presented at the epochs when the LoS of the observations from a given pulsar crosses the bulk of the interplanetary medium close to the Sun, that is, when the SW contribution to DM becomes clear. Figure~\ref{fig:results} presents results from pulsar J0034-0534, during years 2016, 2017 and 2018 and from pulsar J1022+1001 during the years 2016 and 2018. Clearly seen in each panel is a maximum of the DM when the LoS crosses nearest the Sun. All LoSs for pulsar J0034-0534 for the IPS analyses are complete throughout the maximum for the year because no LoS segment comes closer than 15 Rs from the Sun. For pulsar J1022+1001, however, a few LoS segments from 23 August to 30 August come closer than 15 Rs to the Sun (where LoS density are greatest), and thus at these times we have needed to lower the limit of the source surface to 2 Rs in order to complete a representative measurement of Solar proton density. To do this abrogates many of the approximations discussed in the introduction for assumptions made in the 3-D reconstruction analyses. These assumptions are namely that SW densities only obey mass and mass flux conservation (no acceleration), that the observed SW structures move radially outward from the Sun, and that the approximations of Equation~\ref{eq:gleveltomo} are a valid approximation at the LoS close distances from the Sun. All three of these assumptions, and especially the first, are known to be poor approximations of reality. For year 2016 we note that no pulsar dispersion measurements exist for this period when pulsar J1022+1001 was nearest the Sun.

\section{Discussion}\label{sec:discussion}
Overall, these two independent data analyses are in very good agreement. Both show similar increases as the Sun comes close to the LoS of the pulsar. That the IPS 3-D reconstructions show non-uniformity over time attests to the many LoS transient variations present in the analyses from both SIRs and CMEs reconstructed in the data set. Some of these variations appear to match the more direct LoS variations observed in pulsar dispersion measurements. 
However, in this rather sparse data set of pulsar dispersion observations and their comparisons there are a few distinct differences that will require more study, and we would like to discuss these further. There is more variability observed in the pulsar dispersion measurements than in the 3-D reconstructed data; most are positive and somewhat above the smoother 3-D reconstruction result. The values are often greater than the error stated for the pulsar measurements. There is a good chance that many of these differences are real. The 3-D reconstruction result shown at a 6-hour cadence is actually provided by data obtained once a day by the ISEE 327-MHz telescopes which is then smoothed to obtain the best daily result and depicted at 6-hour intervals. The pulsar dispersion measurements are obtained over periods of only about three hours. The heliosphere is known to contain shocks, CMEs, SIR fronts, and other features that provide rapid deviations in the plasma density (see Figure 2 for an example, or \citealt{jbc20} where Thomson-scattering results have been used to reconstruct densities at a cadence of one hour in-situ). If these happen to be present along the LoS near the location most sensitive to their presence, namely closest to the Sun, then they could easily provide an enhancement (or less likely a depletion) relative to the average provided by the 3-D reconstructions. To certify this for the most deviant results would require the pulsar timing be matched as closely as possible with other images of the Sun from Earth - namely coronagraph images, or other remotely-sensed analyses. This work is considered beyond the scope of this current article. 
Beyond this, for both pulsars there appears to be slightly higher measurements for the pulsar data in the regions that probe medium latitudes of the Sun. Whether this is caused by the aforementioned caveat, an incorrect specification of the 3-D LoS extent provided by the 3-D reconstructions that have been extended to only 2.0 AU, an incorrect definition of the conversion of scintillation level to density of Equation~\ref{eq:gleveltomo}, or a poor specification of the background interstellar electron removal from the pulsar dispersion measure will take many more analyses sets and experimentation to discover. Perhaps a slight east-west asymmetry exists whereby the ingress at earlier times has slightly higher values in the 3-D reconstructions than the egress of the pulsar. If so this may be an effect noticed once before for the analyses of data from lower frequency (80 MHz) IPS analysis data from Cambridge, England. That this effect is so slight as to not dominate the results is gratifying for the IPS 327 MHz 3-D reconstruction analyses, as this effect noticed prior was attributed to the elongation of the IPS scintillators along magnetic field lines that are generally smaller to the west of the Sun than to the east. If this were present to a great extent it would mean that it would need to be accounted for by an unknown amount to accurately reconstruct the 3-D data sets. It remains to be determined if this effect exists for the generally lower frequency LOFAR data sets used in these same analyses. That there is a larger mismatch near the Sun for pulsar J0034-0534 for the year 2018 is currently under investigation.

For future comparative analyses of this sort, we would like to obtain more pulsar dispersion analyses from different pulsars over the same time intervals in attempts to provide better comparisons of the variable signals known to exist from transient effects. Additionally, the pulsar data sets can probably be used in the 3-D reconstruction technique to better specify the overall fits of both $\alpha$ and $\beta$ powers of Equation~\ref{eq:gleveltomo}, and if this is the best general relationship to follow. Finally, in a similar analysis it would be good to use this same type analysis close to the Sun and especially at solar minimum in order to probe solar wind acceleration processes in large coronal holes at these locations. These may again help certify some of the solar wind acceleration results determined in the past by other techniques \citep{mj77,swh97,kna98,jyb14}.

\section{Conclusions}\label{sec:conclusions}

In these analyses we have compared the IPS 3-D reconstruction technique with pulsar dispersion measures and found that these give very good comparative results. This opens for a way forward to provide even further studies based on pulsar dispersion analyses, an example of which is reported in Shaifullah et al. (this issue) where the authors validate predictions from the MHD modelling software EUHFORIA (EUropean Heliospheric FORecasting Information Asset, \citealt{pp18}). This bodes well for Solar wind and space weather studies with more data available from either technique.

We have not discussed the results that might be gleaned from comparative rotation analyses of Faraday rotation such studies might bring. The polarization of some pulsar signals are known to exist, but for the heliosphere these are compromised by a much stronger ionospheric signal. The IPS 3-D reconstruction analyses also provide Faraday rotation analyses, but this comes from a combination of the current densities derived and background magnetic fields that lack the most sought (normal) component in the RTN (Radial Tangential Normal) coordinate system. What the 3-D reconstruction analyses can do is provide the R and T (radial and tangential, respectively) components of the background solar wind with density and with those two components times density provide a rotation measure that has the approximate magnitude and variability of the total LOS component in comparison. Perhaps this is a next step these analyses might provide.

\section{Acknowledgments}
The authors wish to thank Dr. Paul Hick, who helped provide much of the original work on the IPS 3-D reconstruction programming is greatly appreciated for his many contributions to these studies. This work is part of the research program Soltrack with project number 016.Veni.192.086 (recipient C. Tiburzi), which is partly financed by the Dutch Research Council (NWO).
C. Tiburzi thanks J. P. W. Verbiest for the useful discussions and comments. 
B.V. Jackson and L. Cota acknowledge funding from NASA contracts 80NSSC17K0684, 80NSSC21K0029, and AFOSR contract FA9550-19-1-0356 to the University of California, San Diego. 
This paper is partially based on data obtained with: i) the German stations of the International LOFAR Telescope (ILT), constructed by ASTRON \citep{vwg13} and operated by the German LOng Wavelength (GLOW) consortium (\url{https://www.glowconsortium.de/}) during station-owners time and proposals LC0\_014, LC1\_048, LC2\_011, LC3\_029, LC4\_025, LT5\_001, LC9\_039, LT10\_014; ii) the LOFAR core, during proposals LC0\_011, DDT0003, LC1\_027, LC1\_042, LC2\_010, LT3\_001, LC4\_004, LT5\_003, LC9\_041, LT10\_004, LPR12\_010; iii) the Swedish station of the ILT during observing proposals carried out from May 2015 to January 2018. We made use of data from the Effelsberg (DE601) LOFAR station funded by the Max-Planck-Gesellschaft; the Unterweilenbach (DE602) LOFAR station funded by the Max-Planck-Institut f\"ur Astrophysik, Garching; the Tautenburg (DE603) LOFAR station funded by the State of Thuringia, supported by the European Union (EFRE) and the Federal Ministry of Education and Research (BMBF) Verbundforschung project D-LOFAR I (grant 05A08ST1); the Potsdam (DE604) LOFAR station funded by the Leibniz-Institut f\"ur Astrophysik, Potsdam; the J\"ulich (DE605) LOFAR station supported by the BMBF Verbundforschung project D-LOFAR I (grant 05A08LJ1); and the Norderstedt (DE609) LOFAR station funded by the BMBF Verbundforschung project D-LOFAR II (grant 05A11LJ1). The observations of the German LOFAR stations were carried out in the stand-alone GLOW mode, which is technically operated and supported by the Max-Planck-Institut f\"ur Radioastronomie, the Forschungszentrum J\"ulich and Bielefeld University. We acknowledge support and operation of the GLOW network, computing and storage facilities by the FZ-J\"ulich, the MPIfR and Bielefeld University and financial support from BMBF D-LOFAR III (grant 05A14PBA) and D-LOFAR IV (grants 05A17PBA and 05A17PC1), and by the states of Nordrhein-Westfalia and Hamburg. CT acknowledges support from Onsala Space Observatory for the provisioning of its facilities/observational support. The Onsala Space Observatory national research infrastructure is funded through Swedish Research Council grant No 2017-00648.

\bibliographystyle{model5-names}
\biboptions{authoryear}
\bibliography{psrrefs,journals,modrefs,crossrefs}

\end{document}